\documentstyle{article}
\begin{document}
\noindent
\large

%\documentstyle[proceedings]{crckapb}
%\begin{opening}
\title{Primal structure of Quantum Mechanics - A critique and an alternative}
\author{S. R. Vatsya}
\date{ }
\maketitle
%\runningtitle{ }
\begin{center}%
%%\\[6ex]
Centre for Research in Earth and Space Science\\[1ex]
York University\\[1ex]
North York, Ontario, Canada~~~~~M3J 1P3\\[1ex]
e-mail: vatsya@cc.UManitoba.CA\\[1ex]
alternate e-mail: oconnorp@acm.org\\[5ex]
\end{center}%
\noindent
\begin{abstract}
The basic premise of Quantum Mechanics, embodied in the doctrine of
wave-particle duality, assigns both, a particle and a wave structure
to the physical entities. The classical laws describing the motion of
a particle and the evolution of a wave are assumed to be correct.
Gauge Mechanics treats the discrete entities as particles, and their
motion is described by an extension of the corresponding classical
laws. Quantum mechanical interpretations of various observations and
their implications, including some issues that are usually ignored,
are presented and compared here with the gauge mechanical
descriptions. The considerations are confined mainly to the conceptual
foundations and the internal consistency of these theories. Although
no major differences between their predictions have yet been noticed,
some deviations are expected, which are indicated. These cases may
provide the testing grounds for further investigations.
\end{abstract}
\section{Introduction}
Early physicists, in their attempts to understand the behaviour of
material bodies in motion, treated them as particles, i.e., as
materially isolated, discrete entities that may be idealized as
mathematical points.
While these structural properties completely characterize a particle,
they give no indication of the laws governing its motion. The motion of a
particle was studied by observing the evolution of its position with
respect to time. Initially, this resulted in the so
called empirical laws, and then in Newton's laws of motion, which were
further adjusted in conformity with the relativistic formulation.
These classical laws were found to describe the motion of particles
encompassing a large range, within the limits of the accuracy of the
measuring devices. However, this success does not establish their
finality, as is the case with all the physical theories. 

In a double slit experiment, photons, electrons and other similar
physical entities encounter a barrier with two slits enabling them to
pass through, and then they land on a distant screen. On the screen
they are known to arrive as isolated, discrete entities with small
extension, that is, as particles, essentially by definition. Now, if \\ \\
A. \ the observed entity is a particle at all space-time points, and \\*
B. \ the classical laws describing the motion of a particle are valid, \\ \\
then a large number of them, arriving as a collection or individually,
should cluster in two separate or overlapping regions on the screen.
However, a large number of the entities identified as particles,
arriving as a collection or individually, are observed to cluster
about several locations, with very few in between, when the separation
between the slits and the screen is sufficiently large [1 pp.2-5].
Thus, if A and B are both true, we encounter a contradiction. Hence, \\ \\
C. \ A is not true or B is not true. \\ \\
To be precise, C holds if one of the following three mutually
exclusive statements is valid:  \\ \\
C1. \ A is not true but B is true. \\*
C2. \ A is true but B is not true. \\*
C3. \ A is not true and B is not true. \\ \\

Quantum Mechanics developed essentially out of the premise C1.,
supplemented with the assumption that these entities are waves at some
space-time points, embodied in the orthodox or the Copenhagen
interpretation. As is well known, various interpretations of Quantum
Mechanics were developed to overcome the consequent difficulties.
Notable among them are the probability interpretations 
[2,3,4 pp.38-44], the pilot wave interpretation, which evolved into Bohmian
Mechanics [5], and the many worlds view [6]. These alternative views
attempt to eliminate C1. as the founding premise of the theory, to
some extent. Also, a number of alternative formulations of mechanics
have been attempted [7] and there are variations on the original
interpretations. 

These attempts were motivated by the fact that a violation of A alone
is sufficient to create some inconsistencies, which are compounded by
the assumption of the wave nature. Thus, a theory based on C3. would
also suffer from such difficulties. Although it is not of primary
importance, C3. is also a stronger statement than either of the other
two. Therefore, it would not be desirable to make C3. the basis of
mechanics, unless a satisfactory theory cannot be based on either of
the others. 

Recently, a new formulation, Gauge Mechanics [8], was developed with
C2. as its premise, supplemented with new laws of motion replacing the
classical ones. The laws of motion for a particle were developed by
extending the classical, by a process of completion in the framework
of Weyl's gauge transformations. Thus, this formulation bypasses the
quantum mechanical and the related developments. The world view
presented by this theory is at variance with the others, and some of
its predictions appear to differ slightly from those of Quantum
Mechanics [8,9]. No major quantitative departures or phenomenological
differences have yet been noticed, and several quantum mechanical
equations have been deduced, as approximations, from this
formulation[8,10].

In the present note, some experimental observations are examined as
interpreted by the quantum and the gauge mechanical formulations. The
focus here is on the logical consistency of their conceptual
foundations. Some of this material has been discussed in great detail
in literature, particularly dealing with Quantum Mechanics, but it is
scattered. We attempt to present a comprehensive picture but
concentrate mainly on the basic concepts, some of which are usually
lost in the details. Also, some relevant points that are frequently
ignored, are discussed here. This comparison shows that Gauge
Mechanics eliminates the troubling aspects of various interpretations
of Quantum Mechanics. Since adequate observations and calculations are
lacking presently, a precise conclusion is not possible but the
estimated deviations in the predictions, whenever they exist, are also
favourable to Gauge Mechanics.

\section{Quantum Mechanics} 
To develop a theory founded on the premise C1.,
a supplementary assumption is needed concerning the structure of these
entities. Violation of A at all space-time points is excluded as it
would contradict the observation of them as particles on the screen,
and also at the source which is established by other observations.
Therefore, a non-particle structure can only be assumed at some 
space-time points.

A classical wave in a double slit experiment is divided in two at the
slits, which interfere with each other as a pair, transmitting energy
to the screen continuously that is distributed in a similar pattern as
the observed density built by the arrival of a large number of the
discrete entities. This similarity provided the basis for the founding
assumption of Quantum Mechanics that what is observed as a particle on
the screen, travels as a wave, which was forged into the doctrine of
the wave-particle duality. The equivalence is established by 
$\lambda = 2\pi/p$, where $\lambda$ is the wavelength of the wave 
and $p$ is the
momentum of the particle, in natural units, i.e., with Planck's
constant equal to $2\pi$.

Even if the history of the formation of the density distribution on
the screen is ignored, the existence of such a pattern is not
sufficient to establish that it was produced by a wave. Therefore, to
have this conjecture as a viable basis of a theory, further evidence
is needed, even if one is inclined to set the following difficulty
aside for the time being. 

Since these entities are observed as particles at some space-time
points, the violation of A requires an abrupt transformation of a non-
particle into a particle on the screen, in fact, whenever an
observation is made. If a compelling evidence of these
entities being waves is found, then this issue could be considered,
likely with some insight provided by the observations. 

The matter of the accuracy of the prediction of this assumption, in
case of the original double slit experiment, has rarely been raised,
but obviously, it is relevant. A coherent wave, split in two at the
slits, must interfere to produce points of zero intensity between the
two consecutive maxima. Observed minimum intensity differs
significantly from zero. Explanations, for example, in terms of the
diffusion caused by the interaction between the observed entities and
the atoms on the screen, are suspiciously qualitative. Thus, to settle
this issue, appears to require more careful measurements.

The predictions based on the rules of Quantum Mechanics are quite
accurate for a large class of phenomena, within the limits of the
measuring devices. In fact, the success of Quantum Mechanical rules in
describing the observations and in the new developments, has been its
strongest defence. However, this does not establish complete accuracy
of the quantum calculations in all situations, and of course the
issue of its conceptual foundations remains open, no matter how
accurate and encompassing the descriptions may be.

Weak conceptual foundations of Quantum Mechanics inspired numerous
experiments, including some thought experiments. These will be stated
and discussed in a later section, on the background of some
interpretations of Quantum Mechanics outlined below.

The original, Copenhagen interpretation is essentially the wave-
particle duality doctrine, that these entities have a dual
personality. The transformation from a wave to a particle is
assumed to be the result of the collapse of the wave-function, caused
by an act of observation, even non-intrusive. This interpretation
introduces unwarranted discontinuities, no mechanism or
characterization for this process has been conceived and it assigns an
unreasonable role to an act of measurement, that of a creator of the
outcome. These are the sources of various difficulties with Quantum
Mechanics [6].  The reverse process, termed the quantum eraser, 
presumably undoes the effect of a collapse [11].

According to the probability interpretation proposed by Schr\"{o}dinger
[2], a physical system has no objective meaning prior to a
measurement, and the associated wavefunction $\psi$, characterizes it
completely. The probability density, given by $|\psi |^2$, is fundamental
and provides a complete description of an individual. The statistical
interpretation suggested by Born [3,4 pp.38-44] is essentially the
same except that it is non-committal about the meaning of a system
prior to a measurement, and thus an individual is described entirely
in terms of a statistical concept. Both of the views hold that a
probabilistic assertion can be a complete description, and retain
discontinuous changes in the wavefunctions as a part of the formalism.
These interpretations attempt to eliminate C1. as the basic premise by
considering the structural properties somewhat irrelevant [12]. 

The pilot wave interpretation, originally proposed by De Broglie and
developed by Bohm [5] is based on the polar representation of the
wavefunction, originally used by Madelung [13]. The wavefunction
$\psi$ may be expressed as $\psi = \sqrt{\sigma} exp(iS)$. Substitution in
the Schr\"{o}dinger equation then yields the following, coupled set of
equations: 
\begin{equation}
\partial \sigma /\partial t + \nabla \cdot (\sigma \nabla S /m) =0 
\end{equation}
\begin{equation}
\partial S/\partial t+(\nabla S)^2/(2m)+V-\nabla^2 \sqrt{\sigma}/
(2m \sqrt{\sigma}) = 0 
\end{equation}

The picture that emerges, from (1), is that the probability density
$\sigma$, evolves as a classical fluid with velocity $(\sigma \nabla S /m)$.
Eq. (2) differs from the Hamilton-Jacobi equation by the last term,
called the quantum potential, that depends on the state of the
particle, and impacts upon the trajectory.

The Madelung equations provided the foundation for what has come to be
known as Bohmian mechanics [5], which is equivalent to Quantum
Mechanics except for a variation in the interpretation. In the
previous interpretations, a particle has no objective meaning or the
issue is irrelevant, and thus, there is no concept of its
trajectories. Instead, the probabilities are considered sufficient to
describe the behaviour of a particle. Bohmian mechanics interprets the
system as a particle following a trajectory, determined by the Bohmian
field, which flows as a classical fluid. Since this is a coupled set
of equations, the trajectory and the field flow, both affect each
other. Thus, Bohmian mechanics interprets Quantum Mechanics so that A
is preserved. The classical law of motion for particles is replaced
by the one determined by the Madelung equations, equivalently, Quantum
Mechanics, involving the Bohmian field, as if a particle floats in it.
In this formulation, particle and wave are made available at all times
acting according to their respective classical characters. Whichever
is suitable for a given situation may then be invoked without forcing
a transformation of one into the other. In case of the double slit
experiment, according to this view, the particle passes through one
slit or the other and the wave, through both, and the probability of a
particle existing in a space-time region is determined by the wave.

Dalton [14] has developed a model with the particle floating in the
fluid. This formulation overlaps with the Bohmian view, but there is a
basic difference: The trajectory is determined by a fundamental force
that coordinates the particle motion with the wave, in a limit. 

Everett's "Relative State" formulation [6], commonly known as the many
worlds interpretation, was developed in an attempt to eliminate the
discontinuous change resulting from an act of measurement. This was
done by considering the original system and the measuring device as
the subsystems of a larger system. The measured state of the observed
subsystem depends on the state of the observing subsystem. The view
presented is that a physical system is composed of many states, each
one in a world of its own. Which state it is observed in, depends on
the world the observer enters. The concept of the collapse of a
wavefunction is avoided by introducing many copies of the world where
a laboratory and the observer exist. All of the other aspects of the
quantum mechanical formulation are maintained. 

All of these formulations maintain the equivalence of the probability
density and $|\psi|^2$, and thus the wavefunctions differing by a phase
factor are physically indistinguishable.

A common feature of all the above interpretations is that each one
addresses a part of the complete set of difficulties associated with
Quantum Mechanics. In some cases, a troubling concept has been
replaced by another, requiring further explanations. For this reason,
the interest in each has varied with time.

\section{Gauge Mechanics} 
The gauge mechanical formulation is
based on the premise C2. from the outset. This requires that the
classical laws of motion for a particle must be modified. The
classical laws describe a large class of phenomena quite accurately.
Therefore, they must be approximations to the more accurate ones.
Thus, their extension should be an appropriate route to the more
accurate laws. Gauge Mechanics is founded on an extension of the
classical, Hamilton's action principle.

So far, the formulation is restricted to the particles of non-zero
rest mass, ignoring the spin. Thus, the following considerations are
strictly applicable only to such cases, although some reasonable
conjectures are possible for the others. An extension to the
particles of zero rest mass appears to be straightforward but the case
of spinors requires additional developments.

\subsection{The gauge mechanical principle}
The concept of the gauge transformations was developed by Weyl,
motivated by the observation that only the ratios, not the absolute
values of the elements of the metric tensor in a space are physically
determinable. Weyl proposed that this makes it impossible to attach an
absolute meaning to the length of a rigid measuring rod. The gauge
group element associated with a general curve is defined in terms of
the change in the rod as it is transported along the curve. The force
field at a point in a space equipped with a gauge field, e.g., the
electromagnetic, may be defined in terms of the change in the length,
as the rod is transported along an infinitesimal closed curve about
the point.

The classical trajectory for a particle is defined by Hamilton's
action principle, i.e., by the requirement that the action about it be
stationary. This is expressed as $\delta S = 0$, up to the first order as
a trajectory is varied, which is equivalent to $exp(\alpha S_{ABA}(\rho_c))=1$ 
up to the first order in the area enclosed by an
arbitrary closed curve $\rho_c$, in a small neighbourhood.  Each curve
$\rho_c$, is defined as the union of a path $\rho'$ from $A$ to $B$, and the
inverted path $\rho$ from $B$ to $A$, and $\alpha$ is a constant. The
gauge mechanical principle extends the action principle by requiring
the equivalence exactly, and includes all the curves into
consideration. Thus, the gauge mechanical, or the physical, paths are
the continuous solutions of
\begin{equation}
\kappa(B) exp(\alpha S_{BA}(\rho)) \kappa^{-1}(A) = 1
\end{equation}
As explained in [8], $\kappa$ represents the state of the system, and
remains constant for a free particle, which may be considered an
extension of Newton's first law. The value of $\alpha$ was also
determined from the motion of the free particles, to be equal to $i$.
It is clear that this approach bypasses the quantum mechanical and the
related developments completely.

The gauge mechanical principle describes motion in terms of the
general gauge group elements while the action principle defines it in
terms of the force field [8]. In addition to the metaphysical
arguments given above, it will be indicated in the next section that
the Aharonov-Bohm effect presents a compelling experimental motivation
for building mechanics in terms of the gauge group elements, instead
of the field. Although of a motivational value, these arguments are
inconsequential for the validity of (3) as the principle describing
the motion of particles. It is sufficient that this prescription
describes the experimental observations in a logically consistent
manner.

Present extension assigns infinitely many, equally likely, paths to a particle 
in motion. A particle may follow any one of the allowed paths on the basis of
random selection. Thus, while there is a definite trajectory that a
particle follows, it cannot be determined before or after the event,
since it is selected on a random basis, even though from a precisely
defined collection, indicating the existence of an objective reality
coupled with randomness, from the outset, which is confirmed by the
later considerations. It should be noted that, there are infinitely many 
trajectories that are excluded from the collection of the particle paths. 

Being an extension, the present approach to the formulation of
mechanics is similar to Hamilton's, which was itself motivated by
Fermat's principle of stationary time for the light rays. All of these
formulations prescribe a principle to assign geometrical trajectories
to the entities under consideration, and thus assume that it can be
idealised as a point, or at least, a central point can be assigned
representing it in a geometrical setting. Einstein's formulation of
the motion of a particle in a gravitational field falls within the
same class. The present approach is geometrical in nature as are
various other theories in physics, including the above, and the modern
ones related to the fields and the fundamental particles.

\subsection{Physical paths} 
The solutions of (3) fall within two distinct
classes: monotonic and non-monotonic, defined by the respective
property of the evolution parameter, which is taken to be the proper
time with the curves being in the Minkowski manifold. The non-
relativistic description is treated as the limiting form of the
relativistic. Along a curve such that all but one coordinate are
constant, a monotonic classical trajectory is physical if and only if
its length is an integral multiple of $2\pi/p$, with p being the
momentum. Each point on the surface of a sphere with radius $2\pi/p$
is reachable along such paths from a point source at the centre. The
next surface is formed as the envelop of the spheres with the same
radius with centres at all points of the original sphere. This
parallels Huygenes' construction for the propagation of a classical
wave. A particle can reach other points also, but along non-classical
physical paths. It was also shown in [8] that a non-monotonic physical
trajectory can be treated as a pair of two monotonic ones, which will
also be called the correlated paths. 

According to some elementary estimates, the physical paths cluster
close to the classical, physical trajectories. As an example, consider
a non-relativistic particle moving along x-axis, from time $t= 0$ to $T$.
Let $\rho$ be a classical trajectory with $x(0) = 0$ and action 
$S(\rho) = (2 \pi n - \omega)$, where $n$ is an integer. 
It is clear that $\rho$ is physical
if and only if $\omega = 0$. It can be shown that there is a physical
path $\hat{\rho}$ in a small neighbourhood of $\rho$ with 
$\rho(0) = \hat{\rho}(0)$ and $\rho(T) = \hat{\rho} (T)$. 
Consider a physical path $\tilde{\rho}$ with $S(\tilde{\rho})= 2 \pi n$, 
in an $\xi$ neighbourhood of $\hat{\rho}$ with $\xi(0) =0$. 
Standard manipulations and estimates yield that 
$\xi(T) \simeq [a \bar{\xi}^2 + b \bar{\xi} \sqrt{\omega}]$, 
where $a$, $b$ are positive constants
and $\bar{\xi}$ is an average value of $\xi$ [9]. The density of such paths
ending close to the end point $\rho (T)$ of $\rho$ is directly proportional
to $[1/\xi(T)] \simeq [a \bar{\xi}^2 + b \bar{\xi}\sqrt{\omega}]^{-1}$. Thus,
there is a considerably high density of allowed paths, and hence of
the particles, about the classical, physical paths than the others.
These estimates are extendable to the case of the non-monotonic
solutions with the same conclusion.

%Note: In the above paragraph, a.e. means almost equal to sign, i.e.,
%one straight bar under a tilde. This material is very close to the
%last two paragraphs of sec. 3. in the last Vigier paper. Equations do
%not have to be displayed, they can be embedded in the text.

\subsection{The double slit experiment} 
%3.3. The double slit Experiment

As discussed above, the quantum mechanical formulation developed from
attempts to reconcile what has been considered the mysterious
behaviour of the microscopic entities in a double slit experiment.
This was done by assigning to each item a dual structural character,
that of a particle and a wave, fused together. Classical laws
describing the time evolutions of the particles and the waves were
assumed to be correct, each one applicable according to the operative
structure of the entity. Gauge Mechanics, on the other hand, considers
the observationally determined structural character definitive, and
the classically unexpected behaviour, indicative of the inadequacy of
the classical laws of motion for a particle. For this to be a viable
theory, therefore, the behaviour of these entities, now treated as
particles, in the double slit experiment should be deducible from the
extended laws. This deduction is outlined below.

In this case, the identical particles encounter two slits, at $A$ and
$A'$, and are collected on a distant screen at a point $B$. Gauge
mechanically, each particle passes through one of the slits. Because
of this arrangement, most of the contribution to the density on the
screen results from the non-monotonic solutions of (3), i.e., the
solutions of 
\begin{equation}
   exp \left[ i(S_{BA}(\rho)-S_{BA'}(\rho')) \right] = \kappa(A')
\kappa^{-1}(A) 
\end{equation}

If $A$ and $A'$ are physically equivalent, $\kappa(A)=\kappa(A')$. From 
the estimates indicated in Sec. 3.2., it follows that most of the
physical paths are concentrated about the trajectories defined by 
\begin{equation}
           (S_{BA}(\rho_s)-S_{BA'}(\rho'_s)) = 2 \pi n 
\end{equation}
where the subscript $s$ indicates that the action is stationary 
and $n$ is an integer.  As a
consequence, the particles cluster about the locations $B$ defined by
(5), in agreement with the observed fact [1 pp.2-5]. With $\Delta r$
being the difference in the path lengths of $\rho_s(AB)$ and
$\rho'_s(A'B)$, (5) reduces to $\Delta r = 2 \pi n/p$.

While there is a qualitative agreement between the gauge mechanical
prediction of the density distribution and that obtained from the
assumption of the wave-particle duality, the quantitative agreement,
although close, is not exact. In particular, it can be shown that
gauge mechanically the minimum density on the screen differs
significantly from zero, in contradistinction with Quantum Mechanics.
However, a close agreement between the two provided the basis for the
deduction of a slightly modified form of Feynman's path integral
formalism of Quantum Mechanics, as an approximation, from the gauge
mechanical formulation [8]. This modification requires that only the
contribution from the physical paths be retained in computing the
probabilities, while Feynman's original formulation retains
contribution from all trajectories. This modification played a crucial
role in the  derivation of some Quantum Mechanical equations,
particularly, the Klein-Gordon equation. Wavefunction in this
derivation appears as a convenient auxiliary quantity. It should be
noted that, while the indicated derivation is instructive, the
original gauge mechanical equations should be used for its more
accurate predictions, which do not involve the wavefunctions.

\section{Discussion}  
In this section we examine the world view presented by
the above formulations by considering some relevant experimental
observations and the related matters. To avoid repetition, various
interpretations of Quantum Mechanics are discussed only when they
provide some additional insight. The world view presented by them is
mostly contained in the comments made in Sec.2., that are easily
applied to most of the phenomenological situations. The gauge
mechanical descriptions are confined to a general exposure that is
sufficient for the present purpose. Further details, particularly the
computational, are available elsewhere, as indicated.

%%\subsection{Wave-particle duality} 
\subsection{The double slit experiment -- Effect of observation}
The conjecture of the wave-particle duality generated an obvious
interest in finding some convincing experimental evidence of these
entities being waves in some space-time regions. The double slit
experiment presents a natural setting for such observations. Whether
an observed entity encounters the slits as a particle or a wave, can
be determined by observing just before it encounters the slits. 
A particle would enter a
slit as a whole, but a wave would divide itself and enter both slits.
To be precise, consider the case of an electron. One may use light to
detect it. If a photon is scattered close to a slit, then the electron
is entering it as a particle. If a reasonably high frequency radiation
is used in this experiment, then an electron is found to enter one
slit, as a complete particle. After a large number of the electrons
have reached the screen, they are found to cluster in two overlapping
regions, which is radically different from the case when they were not
watched. That is, if the electrons are watched, they
demonstrate particle structure and travel according to the classical
laws of motion for a particle: both $A$ and $B$ appear to be the correct
assumptions.

This experiment may be performed with lower frequency radiation to
reduce the impact of the observation on the electron. If the frequency
is very low, the interference like distribution on the screen is
maintained. At these frequencies, one cannot determine which slit the
electron passes through. To be consistent with Quantum Mechanics, a
photon cannot be localized with better precision than its wavelength.
At a low frequency, the photon may have scattered at any point in a
region including both slits. Thus, if the electrons cannot be
determined to have passed through one of the slits as particles, 
they appear to pass through both slits as waves.

The density distributions on the screen, as the frequency of the
observing photons is altered gradually, appears to be unavailable.
According to the wave-particle duality, the distribution should change
little as the frequency is reduced, and change abruptly to an
interference like pattern as the wavelength becomes large enough to
cover both slits. That is, there is a critical level of intrusion to
cause the collapse of the wave. Bohmian mechanics avoids the concept
of collapse, but the particle gets disentangled from the Bohmian field
in a discontinuous manner. In the many worlds view, which world the
observing device enters depends discontinuously on the level of
intrusion it creates. 

As indicated in Sec. 3.3., gauge mechanically, the existence of an
interference like pattern requires that 
$\delta \kappa = (\kappa (A) - \kappa (A')) = 0$. 
The solutions of (4) depend continuously on $\delta\kappa$, 
and thus, the pattern on the screen should change continuously
with its value. It can be shown that for sufficiently large value of
$|\delta \kappa|$, most of the contribution to the density on the screen
results from the monotonic solutions of (3), which yields a classical
particle like distribution [8,9]. An intrusive observation alters the
value of $\delta \kappa$ which can be determined or estimated with the
same degree of accuracy as the available details of the interaction
[9]. In the above experiment, the pattern on the screen should change
continuously with respect to the momentum imparted by the colliding
photon to the electron, from one type to the other.

\subsection{The uncertainty principle} 
The uncertainty principle may be
deduced from the above observation and the assumption of the wave-
particle duality, as follows. The bulk of the argument, briefly
outlined here, is the same as in ref. 1, pp.9-13.

Since the point of arrival of an individual particle on the screen
does not enable one to determine which slit it may have passed
through, its momentum p has an uncertainty of $\delta p$, given by
$(\delta p/p) \simeq (\delta r/d)$, where $d$ is the separation
between the two consecutive maxima on the screen. Here, 
$\delta r = \Delta r$ for $n=1$ as defined in Sec. 3.3. 
A determination of $p$ and
position with an accuracy of $\delta p$ and $d/2$ respectively, would
enable one to construct the distribution that is observed for a
collection of the undisturbed particles, which is prohibited by the
observed behaviour of the electrons when watched with photons with
this precision. Hence, $\delta p \delta x \geq p\delta r/2$.

In Quantum Mechanics, the wave-particle duality associates a wave of 
wavelength $\lambda = 2\pi/p$, with a particle with momentum $p$. 
The relation $\delta r = \lambda$ follows from the classical laws of 
wave propagation. Both relations together yield $\delta r = 2\pi/p$. Hence, 
$\delta p \delta x \geq \pi$. 

The interpretations of Quantum Mechanics assume the uncertainty
principle as a part of their schemes. In Gauge Mechanics, the relation
$\delta r = 2\pi/p$ follows from (5). Thus, the uncertainty principle
follows without invoking any concepts other than the gauge mechanical,
and without any reference to the experimental observations. The
behaviour of the particles in the double slit experiment is 
deducible from the basic formulation of Gauge Mechanics. Therefore, the
uncertainty principle also is a deduced result.

\subsection{The double slit experiment -- Delayed choice observation}
The impact of an intrusive observation on the behaviour of the
electrons, before they enter the slits, can be eliminated completely,
by watching them between the slits and the screen, i.e., the delayed
choice experiment [11]. Again, the electrons are found to pass through
one slit or the other, as complete units, and appear to travel
according to the classical laws of motion for a particle, if high
frequency photons are used to watch. 
With low frequency photons,
they appear to act as waves. The comments concerning the use of 
intermediate frequency photons, made in Sec. 4.1, apply to this case also.
The quantum mechanical understanding of this behaviour creates a
paradoxical situation: Each electron knew in advance how it would be
observed. 

Gauge mechanically, the situation in a delayed choice experiment is
identical to 4.1.: A particle passes through one slit or the other.
The momentum imparted during an intrusive measurement alters its
course in a quantifiable way to result in the expected observation [9].

Incidentally, this behaviour, instead of Sec. 4.1, may be invoked in
the deduction of the uncertainty principle in Sec. 4.2 [1 p. 7-13].

\subsection{The double slit experiment -- Effect of screen's location}
In the original double slit experiment, a large number of
electrons produce an interference like pattern on a distant screen. If
the screen is moved close to the slits, one finds the electrons
clustered about two separate locations. 

Implications of this observation appear to be minimised in literature
by arguing that moving the screen closer to the slits forces an
electron to choose between the slits [11], and thus, to behave as
a particle. There is no difference between this setup and the same
setup with a larger separation. If a particle passes through both
slits as a wave, the pattern should be interference-like, which would
give no indication of which slit the electron passed through. Thus, no
agency has forced an electron to behave as a particle and no
additional information has been generated by the experimental setup
that would enable one to
determine which slit the electron passed through. 
This information is deducible only from the density distribution on the screen.

Gauge mechanically, it is clear that the paths available to a particle
are determined by the experimental setup. Most of the particles
passing through each slit would cluster close to the monotonic,
classical physical paths. If the screen is close to the slits so that
the regions containing the end points of these paths do not overlap,
this covers most of the paths. Usual estimates then yield the observed
classical particle like behaviour. If these regions overlap, there is
a host of correlated paths available, which outnumber the monotonic
trajectories. As in Sec. 3.3., this yields an interference like
pattern. According to the estimates, the observation on the screen
should change gradually as the screen is moved closer to the slits.

\subsection{The Aharonov-Bohm effect [15]} 
In the experimental setup for this
case, identical electrons travel from $A$ to $C$ to $B$, and from $A$ to $D$ to
$B$, shielded from the magnetic field that the closed path
$\rho_c(ACBDA)$ encloses. Here $B$ is a point on the screen. As in the case
of the double slit experiment, an interference like pattern forms on
the screen. This pattern changes continuously with the magnetic flux $F$, 
repeating the original one for each integer $n$ as 
$F = \oint_{\rho_c(ACBDA)} \phi_\mu dx^\mu$, 
is replaced by $(F+2 \pi n)$. Thus, the pattern changes
continuously as the electro-magnetic potential changes from $\phi$ to
$\phi'$, repeating itself whenever
\begin{equation}
exp[i \oint_{\rho_c(ACBDA)} (\phi_\mu - \phi'_\mu) dx^\mu] = 1
\end{equation}
i.e., whenever the gauge group elements assigned by them to the curve
$\rho_c(ACBDA)$ are equivalent. 

All interpretations of Quantum Mechanics equate $|\psi|^{2}$ with the
probability density, which determines all the physically observable
quantities. This implies an equivalence of all wavefunctions differing
by a phase factor. The Aharonov-Bohm effect shows that the two
wavefunctions differing by a phase factor produce a physically
measurable effect. This effect was predicted on the basis of the
quantum mechanical equations, indicating an incompatibility of the
equations with one of its basic assumptions. 

Two alternatives have been suggested to reconcile this experimental
observation with the quantum mechanical formulation. One suggestion is
to re-interpret the wavefunction and the other, to assume action at a
distance. Both of these alternatives create additional difficulties.
Also, it raises the issue whether electro-magnetism and other forces,
are adequately described by the fields or by the potentials. The view
that emerges, from (6), is that the field under-describes and the potential 
over-describes them. A complete and optimal description is provided by the
phase-factors, i.e., the gauge group elements [16].

This experiment could be conducted without any reference to Quantum
Mechanics with the same observation, which alone is sufficient to
provide a strong motivation for developing a theory of mechanics that
would describe the motion of a charged particle in terms of the gauge
group elements. Although arrived at by some metaphysical reasoning,
the gauge mechanical principle formulates mechanics in terms of the
gauge group elements from the outset. Thus the original arguments
support and are supported by this observation.

The gauge mechanical description of this effect is straightforward
[8,9]. The major facilitators for the passage of the particles in this
case are the solutions of
\begin{equation}
     exp[i(S_{ADBCA}^P (\rho_c) - F)] = 1 
\end{equation}
where  $S_{ADBCA}^P (\rho_c)$ is the free particle part of the action. 
As in Sec.3.3., this implies a density pattern similar to the case of
the double slit experiment, changing continuously with $F$, repeating
the original pattern for each integer n as $F$ is replaced by 
$(F+2 \pi n)$, in agreement with the experimental observation [15].

\subsection{The EPR problem}

Einstein, Podolsky and Rosen devised a thought experiment to argue
that Quantum Mechanics is an incomplete and inadequate theory [17]. In
the experiment, two particles in a bound state break up and travel in
opposite directions. The pair is correlated, e.g., both must have
equal and opposite momenta. Their relative position is also conserved.
An intrusive measurement may be made on one of the particles to
determine its momentum with complete accuracy by leaving its position
completely undetermined. This determines the momentum of the other
particle with complete precision, by the conservation law. Since the
measurement can be made in time less than needed for light to travel from 
one particle to the other, the authors argued that the momentum
of the undisturbed particle must have been defined from the beginning.
Since Quantum Mechanics is unable to assign this value, it is
incomplete and fundamentally inadequate.

Also, the position of the second particle can be determined with
complete precision, as a complete indeterminancy in its momentum can
be tolerated, which has already been determined exactly. This also
determines the exact position of the first particle as the relative
position of the two is conserved. Thus, the positions and momenta of
both particles are precisely determined, in violation of the
uncertainty principle.

Above argument assumes that the information from one particle to the
other cannot travel faster than light. Therefore, an alternative to
the above scenario is the possibility of passage of signals faster
than light, in fact instantaneous, i.e., action at a distance. This 
non-locality was unacceptable to Einstein. 

Above arguments are applicable, not just to the position and momentum,
but also to various other physical quantities. There are several
alternative measurements that can be made to check if the non-local
effects exist. Bell's inequalities [18] greatly facilitated such tests
[19], which have been conducted and non-locality confirmed [see e.g.
20]. Thus, non-locality is an experimental fact, but it can be
reconciled with the relativistic assumption of the limiting speed. If
the momenta and the positions of the equivalent particles are
measured, they are found to be randomly distributed. Thus, randomness
is also an experimental fact. A faster than light transmission 
between a pair of correlated particles is possible but it
is only the randomness that would be so transmitted. This makes it
impossible for a precise signal to travel faster than light. 

Such non-local effects are inherent in Quantum Mechanics. In the gauge
mechanical terms, this arrangement is essentially equivalent to the
double slit experiment. Classical physical paths $\rho(AB)$ and
$\rho'(AC)$, are defined by
\begin{equation}
\kappa(B)exp[iS_{BA}(\rho)]exp[-iS_{CA}(\rho')]\kappa^{-1}(C)=1 
\end{equation}
If the particles travel undisturbed, i.e., 
$\kappa(A) = \kappa(B) = \kappa(C)$, then the solutions are given by 
\begin{equation}
             exp[i(S_{BA}(\rho)-S_{CA}(\rho'))] = 1 
\end{equation}
If the state of one of the particles is altered by an intrusive
measurement, say at $B$, then $\kappa(B) \neq \kappa(A) = \kappa(C)$,
and hence, the solutions of (8) no longer satisfy (9). 
The corresponding physical
path now is the union of $\rho(AB)$ and $\rho''(AC') \neq \rho'(AC)$,
compensating for the state change by a multiplicative factor equal to
$exp(i \delta S)$ [21], where $\delta S$ is the measure of the
intrusion.

Gauge mechanically, the positions and the momenta of both of the
correlated particles are precisely defined, before and after a
measurement. A change occurs as a result of an intrusion which is
quantifiable, that is communicated to the undisturbed particle. Thus
non-locality is a gauge mechanically predicted effect. 

\subsection{Potential barrier}

According to the quantum mechanical view [22], if a particle
encounters a classically forbidden barrier, the wave-packet divides
itself in two. One part is reflected and the other, passes through.
Thus, out of a large collection, most particles are reflected but some
tunnel through. As soon as a particle is detected, both parts of the
wave-packet disappear, or become inconsequential, depending on the
interpretation used. The gauge mechanical description [9] is outlined
below.

For a relativistic particle, the action along a geodesic line is given
by
\begin{equation}
S_{BA}(\rho_c) = - m \sqrt{t(l)^{2} - l^{2}} - V t(l)  
\end{equation}
where $V$, the potential, is zero outside the barrier, and
positive inside. Here, $l$ is the length of the line segment and $t(l)$
is the classical time taken by the particle in travelling the distance
$l$. For a piecewise geodetic trajectory, the action is given by the
sum of the actions on each of the segments. Incidentally, the 
non-relativistic approximation to this value, with $V=0$, was used to
obtain the estimates of Sec. 3.2.

As in Sec.3.2., most of the physical paths are concentrated about the
piecewise classical, physical paths. Thus, most of the particles would
reflect from a classically forbidden barrier. It is straightforward to
check, by direct substitution, that there are physical paths allowing
tunnelling with emerging particles having a large spread of
velocities, but not exceeding the speed of light [9].

\subsection{Schr\"{o}dinger's cat[11,23]}

In this thought experiment, a cat is in a box together with a lump of
radio-active material, which may decay releasing a particle that may
trigger a hammer, smashing a vial containing cyanide gas, and killing
the cat. According to the Copenhagen interpretation, the cat is both
dead and alive until an observation is made, which collapses it in one
of the two states. The situation is complicated by the fact that this
observation can be non-intrusive. The interpretations that assign a
probability to an event are meaningful only if many cats are observed.
The many worlds interpretation determines the cat to be dead in one
world and alive in the other. What an observer would find depends on
which world one enters, which is indeterminable until after the
outcome of the observation. In the limit of infinite time, then there
is no choice for any observer but to enter the world of the dead cat.

Gauge mechanically, as discussed in Sec. 4.7., there are some paths
connecting an interior point of the radio-active material to a point
outside, facilitating the passage of a particle to the trigger. By a
given time, if a particle has triggered the hammer, the cat is dead,
otherwise it is alive. Thus, an objective reality exists, i.e., the
cat is either dead or alive, but because of randomness, one cannot
determine the state without an observation, which however, has no
impact on the state, on the objective reality, only determines it as
it exists.

\section{Concluding Remarks}

The underlying theme in the quantum mechanical formulation of dynamics
is that an observationally isolated discrete entity of limited
extension, is a self-interacting continuous system of infinite
extension, resulting from something oscillating, when not observed.
Various paradoxical situations result from this underpinning. The so
called quantum mysteries result from the attempts to understand the
physical phenomena in terms of the classical concepts: particles
travelling according to the classical laws and waves described by the
oscillations of the particles, or fields. Objections to the claim of
completeness of Quantum Mechanics emanate from its treatment of a
single system in terms of the concepts that are pertinent to a
statistical collection. A number of interpretations have been
developed to reconcile the inconsistencies, e.g., the Copenhagen, the
probability, the many worlds and the pilot wave, that evolved into
Bohmian mechanics. These interpretations and their limitations are
well known. In brief, each one addresses only a part of the
difficulties. At times the explanations generate new questions.

Gauge Mechanics is founded upon an extension of the classical laws
governing the motion of the localized entities, the particles, without
being prejudiced by the quantum mechanical thinking. Since the
structural properties of the particles remain intact, the related
observations are described in a consistent manner. The extension
determines a collection of equally likely trajectories for a particle
to follow. Which one of the paths is followed by a particular particle
is determined on the basis of random selection. Consequently, each
particle has definite properties, e.g., the momentum and the position,
but a quantitative plot representing the collection must show a
spread, resembling the envelop of a wave packet or something related.
An act of an intrusive observation may alter the properties of a
particle which is understandable in terms of a physical act of
objective and quantifiable nature, but a non-intrusive one only
delineates the reality. 

Although Gauge Mechanics assigns a probability to an event, it arises
out of the statistical behaviour of an ensemble of paths. 
The randomness in its original formulation is a
consequence of the existence of a collection of solution trajectories
of the basic equation. A random behaviour of the identical particles
is an experimentally observed fact. The non-local effects, which have
been experimentally verified, are also described by the gauge
mechanical formulation but the view differs fundamentally from the
quantum mechanical.

The estimates obtained so far indicate that while there are
differences between the quantum mechanical and the corresponding gauge
mechanical predictions, they must be small. A more accurate evaluation
of the gauge mechanical probabilities requires more intricate
analysis. However, satisfactory approximations may be obtained by
straightforward numerical computations. Further, a better comparison
with the observations requires more careful measurements. A measurable
difference, if found, should determine which one of the two theories
describes the experimental observations more accurately. 

It is clear also, that further analysis is needed for a more complete
understanding of the implications of the gauge mechanical formulation.
If this theory is found to be more satisfactory than the existing
Quantum Mechanics in describing the motion of particles, then the
classical laws governing the structure and evolution of waves, and
fields in general, must also be examined and adjusted accordingly, if
need be.

\begin{center}
{\bf Acknowledgements}\\*
\end{center}
The author is thankful to Professor Huw O. Pritchard for hospitality,
and to Patrick A. O'Connor for discussions and technical help. \\[3ex]

References:

\begin{enumerate}
\item{}
R.P. Feynman and A.R. Hibbs, Quantum Mechanics and Path Integrals,
McGraw-Hill, New York, 1965.
\item{}
E. Mackinnon, in Studies in the Foundations of Quantum Mechanics,
edited by P. Suppes (Philosophy of Science Association, East Lansing,
MI, 1980) pp. 1-56
\item{}
M. Born, Science 122 (1955) 675-679; A. Pais, Science 218 (1982)
1193-1198
\item{}
M. Jammer, The Philosophy of Quantum Mechanics, Wiley, New York,
1974
\item{}
D. Bohm, Phys. Rev., 85 (1952) 169; D. Bohm and J.P. Vigier, Phys.
Rev., 96 (1954) 208.
\item{}
H. Everett, Rev. Mod. Phys., 29 (1957) 454; J.A. Wheeler, Rev.
Mod. Phys., 29 (1957) 463.
\item{}
D. Bohm and B.J. Hiley, The Undivided Universe, Routhledge, London
and New York, 1993; P.R. Holland, The Quantum Theory of Motion,
Cambridge University Press, 1993; Quantum Mechanics and Chaotic
Fractals, Special Issue of Chaos, Solitons \& Fractals 4(3) (1994)
(Edited by M.S. El Naschie and O.E. Rossler) 
\item{}
S.R. Vatsya, Can. J. Phys. 73 (1995) 85; S.R. Vatsya, Can. J.
Phys., 67, (1989) 634.
\item{}
S.R. Vatsya,In Causality and Locality in Modern Physics (ed. G.
Hunter, S. Jeffers and J.P. Vigier, 1998) Kluwer Academic Publishers,
Dordrecht, pp.243-251.
\item{}
S.R. Vatsya, Can. J. Phys., 73, (1995) 602, gr-qc/9412004 ;  
S.R. Vatsya, Chaos, Solitons \& Fractals, 1998, In press.
\item{}
J. Horgan, Quantum Philosophy, Scientific American, July 1992,
pp.94-104.
\item{}
H. Feshbach and V.F. Weisskopf, Phys. Today 41 (1988) 9
\item{}
E. Madelung, Z. Phys. 40 (1926) 322.
\item{}
B.J. Dalton, Deterministic Explanation of Quantum Mechanics Based
on a New Trajectory-Wave Ordering Interaction, North Star Press of St.
Cloud, Inc., St. Cloud, Minnesota, 1994; In. J. Theor. Phys. 21 (1982)
765
\item{}
Y. Aharonov and D. Bohm, Phys. Rev. 115 (1959) 485; A. Tonomura,
et al, Phys. Rev. Lett. 56 (1986) 792; M.P. Silverman, Am. J. Phys. 61
(1993) 514.
\item{}
T.S. Wu and C.N. Yang, Phys. Rev. D 12 (1975) 3845
\item{}
A. Einstein, B. Podolsky and N. Rosen, Phys. Rev. 47 (1935) 777;
N.D. Mermin, Physics Today, April 1985, pp. 38-47.
\item{}
J.S.Bell, Physics 1 (1965) 195
\item{}
J.F. Clauser and A. Shimony, Rep. Prog. Phys. 41 (1978) 1881. 
\item{}
A. Aspect, P. Grangier and G. Roger, Phys. Rev. Lett. 47 (1981)
460; A. Aspect, P. Grangier and G. Roger, Phys. Rev. Lett. 49 (1982)
91; A aspect, J. Dalibard and G. Roger, Phys. Rev. Lett 49 (1982) 1804
\item{}
S.R. Vatsya, in "The Present Status of the Quantum Theory of
Light" (ed. S. Jeffers et al), Kluwer Academic Publishers,
Netherlands, 1997, pp. 223-234; quant-ph/9601003
\item{}
R. Chiao, P.G. Kwiat and A.M. Steinberg, Scientific American,
August 1993, p. 52.
\item{}
E. Schr\"{o}dinger, Naturwissenschaften, 23 (1935) 807

\end{enumerate}

\end{document}